\documentclass[preprint]{aastex}

\usepackage{lmacs}

\received{} 
\accepted{} 
\journalid{}{} 
\articleid{}{} 
\shortauthors{Baron et al.}
\shorttitle{Spectral Analysis of  SN 1999\lowercase{em}}

\begin{document}

\title{Preliminary Spectral Analysis of the Type II 
Supernova 1999\lowercase{em}}

\author{{E.~Baron\altaffilmark{1}},
{David~Branch\altaffilmark{1}}, {Peter~H. Hauschildt\altaffilmark{2}},
{Alexei~V.~Filippenko\altaffilmark{3}}, {R.~P.~Kirshner\altaffilmark{4}},
{P.~M.~Challis\altaffilmark{4}}, {S.~Jha\altaffilmark{4}},
{R. Chevalier\altaffilmark{5}},  
 {Claes Fransson\altaffilmark{6}},
{Peter Lundqvist\altaffilmark{6}}, {Peter Garnavich\altaffilmark{7}}, 
{Bruno Leibundgut\altaffilmark{8}}, {R. McCray\altaffilmark{9}}, 
{E. Michael \altaffilmark{9}},
{Nino Panagia\altaffilmark{10}}, {M.~M.~Phillips\altaffilmark{11}},
{C.~S.~J.~Pun\altaffilmark{12}}, {Brian Schmidt\altaffilmark{13}},
{George Sonneborn\altaffilmark{12}}, 
{N.~B.~Suntzeff\altaffilmark{14}}, 
{L.~Wang\altaffilmark{15}} and {J.~C.~Wheeler\altaffilmark{15}}}

\altaffiltext{1}{Department of Physics and Astronomy, University of
Oklahoma, Norman, OK~73019}

\altaffiltext{2}{Department of Physics and Astronomy \& Center for
Simulational Physics, 
University of Georgia, Athens, GA 30602}

\altaffiltext{3}{Department of Astronomy, University of California,
Berkeley, CA~94720--3411}

\altaffiltext{4}{Harvard--Smithsonian Center for Astrophysics,
60~Garden St., Cambridge, MA~02138}

\altaffiltext{5}{Dept. of Astronomy, Univ. of Virginia, P.O. Box 3818,
Charlottesville, VA 22903}

\altaffiltext{6}{Stockholm Observatory, SE--133~36 Saltsj\"obaden,
Sweden}

\altaffiltext{7}{Dept. of Physics, Univ. of Notre Dame, 225 Nieuwland
Science Hall, Notre Dame, IN 45656}

\altaffiltext{8}{European Southern Observatory,
Karl-Schwarzschild-Strasse 2, D-85748 Garching, Germany}

\altaffiltext{9}{JILA,  Univ. of Colorado, 
Boulder, CO 80309}

\altaffiltext{10}{Space Telescope Science Institute, 3700 San Martin
Drive, Baltimore, MD~21218 (on assignment from Space Science
Department of ESA)}

\altaffiltext{11}{Carnegie Inst. of Washington,
                        Las Campanas Obs.,
                        Casilla 601, Chile}

\altaffiltext{12}{Laboratory for Astronomy and Solar Physics,
NASA/GSFC, Code~681, Greenbelt, MD~20771}

\altaffiltext{13}{Mount Stromlo Obs, Australian National Univ.
			Private Bag,
			Weston Creek P.O, ACT 2611, Australia}

\altaffiltext{14}{CTIO, NOAO, Casilla~603, La~Serena, Chile}

\altaffiltext{15}{Department of Astronomy, University of Texas,
Austin, TX~78712}

\begin{abstract}
We have calculated fast direct spectral model fits to two early-time
spectra of the Type-II plateau SN 1999em, using the SYNOW synthetic
spectrum code. The first is an extremely early blue optical spectrum
and the second a combined \emph{HST} and optical spectrum obtained one
week later.  Spectroscopically this supernova appears to be a normal
Type II and these fits are in excellent agreement with the observed
spectra. Our direct analysis suggests the presence of enhanced
nitrogen. We have further studied these spectra with the full NLTE
general model atmosphere code \texttt{PHOENIX}. While we do not find
confirmation for enhanced nitrogen (nor do we rule it out), we do
require enhanced helium. An even more intriguing possible line
identification is complicated Balmer and He~I lines, which we show
falls naturally out of the detailed calculations with a shallow
density gradient. We also show that very early spectra such as those
presented here combined with sophisticated spectral modeling allows an
independent estimate of the total reddening to the supernova, since
when the spectrum is very blue, dereddening leads to changes in the
blue flux that cannot be reproduced by altering the ``temperature'' of
the emitted radiation. These results are extremely encouraging since
they imply that detailed modeling of early spectra can shed light on
both the abundances and total extinction of SNe II, the latter
improving their utility and reliability as distance indicators.
\end{abstract}


\section{Introduction}

SN 1999em was discovered by the Lick Observatory Supernova Search
(IAUC 7294) on Oct 29.44 UT and confirmed by the BAO supernova group on
Oct 29.7 UT. It was followed spectroscopically and photometrically by
ground-based observatories and was the subject of a Supernova
Intensive Studies (SINS) \emph{Hubble Space Telescope} (\emph{HST})
observation on Nov 5, 1999.  
SN 1999em is in the SABc galaxy NGC 1637, with a heliocentric
recession velocity of 717 \kmps. The first spectrum taken on Oct 29,
1999 is extremely blue and displays strong Balmer features typical of
Type II supernovae with a characteristic velocity of around 10,000 \kmps.

SN 1999em appears to have been a normal Type IIP, likely the result of
a core-collapse in a star with a massive hydrogen envelope
\citep{leon99em00}. The early and weak radio emission (IAU Circulars
7318 and 7336) hints that the star did not experience significant mass
loss shortly before the explosion, which is also thought to be the
normal situation for SNe IIP \citep[see][and references
therein]{weilermd00}. The lack of circumstellar interaction makes SNe
IIP ``clean'' cases to model, resulting in more reliable atmospheric
modeling which in turn makes SNe IIP ideal for cosmological
use. Because SN 1999em was observed earlier, and over a wider
wavelength range than any other normal bright SN IIP, it is a perfect
test bed for our understanding of SNe IIP.  These supernovae are of
particular interest as distance indicators.

\section{Observations}

The observed spectra we use in this paper were obtained on
Nov 4, 1999 at the Fred L. Whipple Observatory (FLWO), and on
Oct. 29, 1999 at CTIO. \citet{leon99em00} find that on Oct. 29 the
\emph{V} light  
curve is rising. The point at
Nov.~4-5 is on a broad maximium prior to the
plateau. The light curve settles down to a plateau phase on about
Nov.~20.

\section{SYNOW Models}

We have used the fast, parameterized supernova synthetic spectra code
SYNOW to 
make an initial investigation of line identifications and expansion
velocities.  The code is discussed in detail by \citet{fisher00} and
recent applications include \citet{fisher91T99}, \citet{millard94i99},
 and \citet{hatano94D99}.  SYNOW spectra consist of
resonant--scattering profiles superimposed on a blackbody continuum.
For the two synthetic spectra presented here, the radial dependence of all
line optical depths is a power law of index 8.  The observed spectra
have been corrected for a redshift of 717 km~s$^{-1}$ and for
reddening of $E(B-V)=0.05$~mag (see below).

In Figure~\ref{syn_oct29} the CTIO observed spectrum of October~29 is
compared with a synthetic spectrum for which the continuum blackbody
temperature $T_{bb}$ and the excitation temperature $T_{exc}$ are
13,000~K, the velocity at the photosphere $V_{phot}$ is 11,000
km~s$^{-1}$, and only lines of hydrogen, He~I, and N~II are
considered.  The relative strengths of the lines of an ion are fixed
by local thermodynamic equilibrium (LTE) at the excitation
temperature, but the absolute strengths of the lines of each ion are
controlled by a free parameter.  The strength of the hydrogen lines
has been chosen to make H$\beta$ and H$\gamma$ about the same as in
the observed spectrum; the relative optical depths of the hydrogen
lines are of course fixed by atomic physics so the poor fit to
H$\alpha$ is due to a deficiency of the resonant--scattering source
function for this transition.  The only noticeable line of He~I in the
synthetic spectrum is that of $\lambda$5876.  Lines of N~II have been
introduced so that $\lambda4623$ and $\lambda5679$ can account for the
absorptions near 4500 and 5500~\AA, in which case N~II $\lambda5029$
also affects the synthetic spectrum slightly near 5000~\AA.  (In SYNOW
when a particular species in introduced, all the lines of that species
are included with their relative strengths set by assuming LTE
populations, but we focus here on the strong optical features which
are relevant to the observed spectra.) The hydrogen and He~I
identifications are definite, and (in anything remotely like LTE) we
can offer no alternative to the N~II lines.  However, see
\S~\ref{complicated} regarding the possibility that some of the
``N~II'' lines may actually be due to hydrogen.

\begin{figure} 
\begin{center}
\leavevmode
\epsscale{0.95} 
\plotone{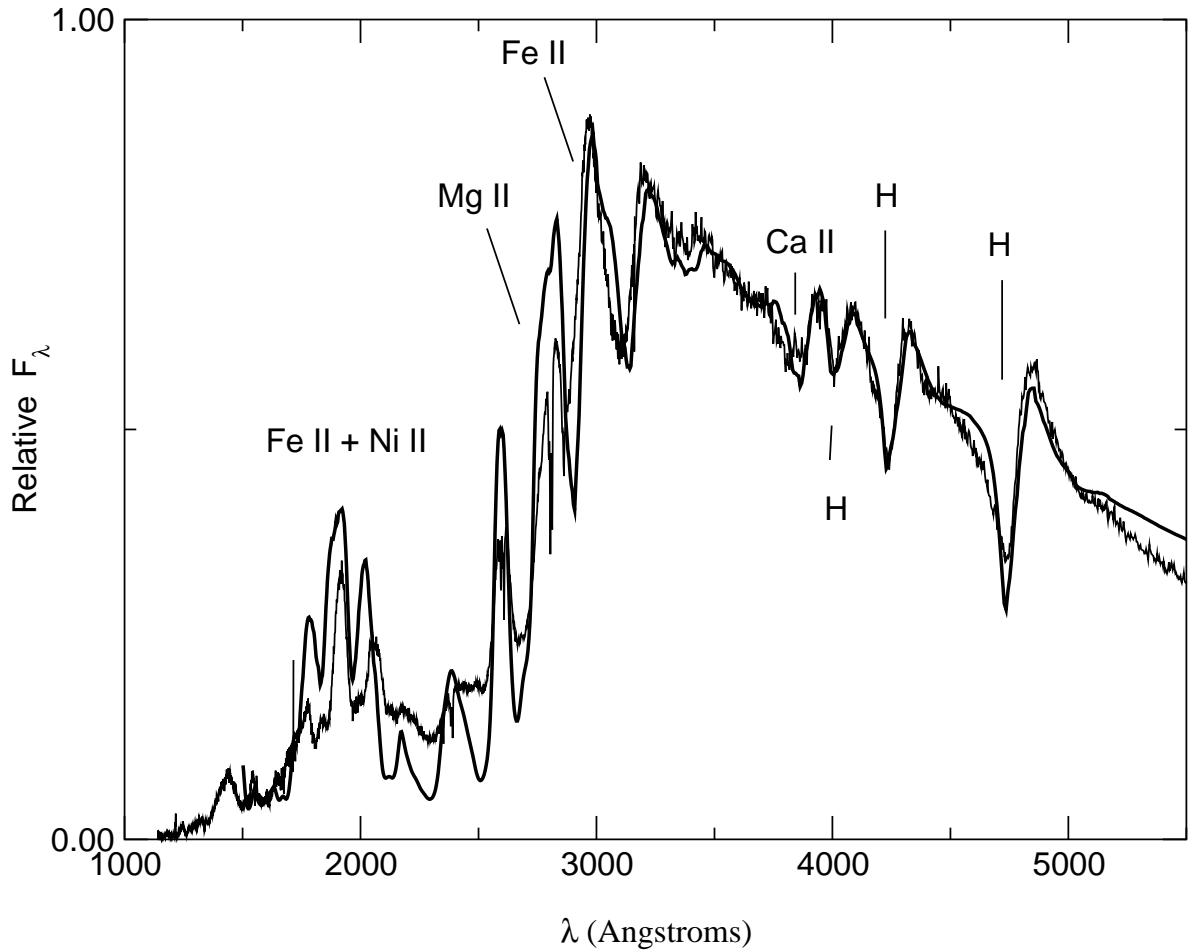}
\caption{\label{syn_nov4} SYNOW fit to the UV+optical \emph{HST}
spectrum obtained on Nov. 5, with the optical spectrum obtained at the
FLWO on Nov. 4. The \emph{HST} and optical spectra overlap at around
3600~\ang. The observed spectra has been de-reddened using $E(B-V)=0.05$.
The fit quality should be judged primarily in the
optical.}
\end{center}
\end{figure}

In Figure~\ref{syn_nov4} the \emph{HST} observed spectrum of November~5 is
compared with a 
synthetic spectrum that has $T_{bb} = T_{exc} = 9500$~K, $v_{phot} =
8000$ km~s$^{-1}$, and lines of hydrogen, Ca~II, Mg~II, Fe~II, and
Ni~II.  The resonance--scattering approximation gives
good fits to the features produced by H$\beta$, H$\gamma$, and
H$\delta$.  Ca~II contributes only the blend due to $\lambda$3945 (the
H and K lines) and Mg~II contributes little other than the blend due
to $\lambda2798$ (the h and k lines).  All other features in the
synthetic spectrum are produced by Fe~II and, to a lesser extent, by
Ni~II.  The hydrogen, Ca~II, and Mg~II identifications are definite.
The ultraviolet spectrum evidently is mainly a blend of
singly--ionized iron--peak lines; Fe~II definitely is present and
Ni~II probably contributes significantly.  Other iron--peak ions that
have not been introduced here may also affect the observed spectrum.

\section{PHOENIX Models}

\subsection{Oct 29\label{phxoct29}}

\begin{figure} 
\begin{center}
\leavevmode
\epsscale{0.95} 
\plotone{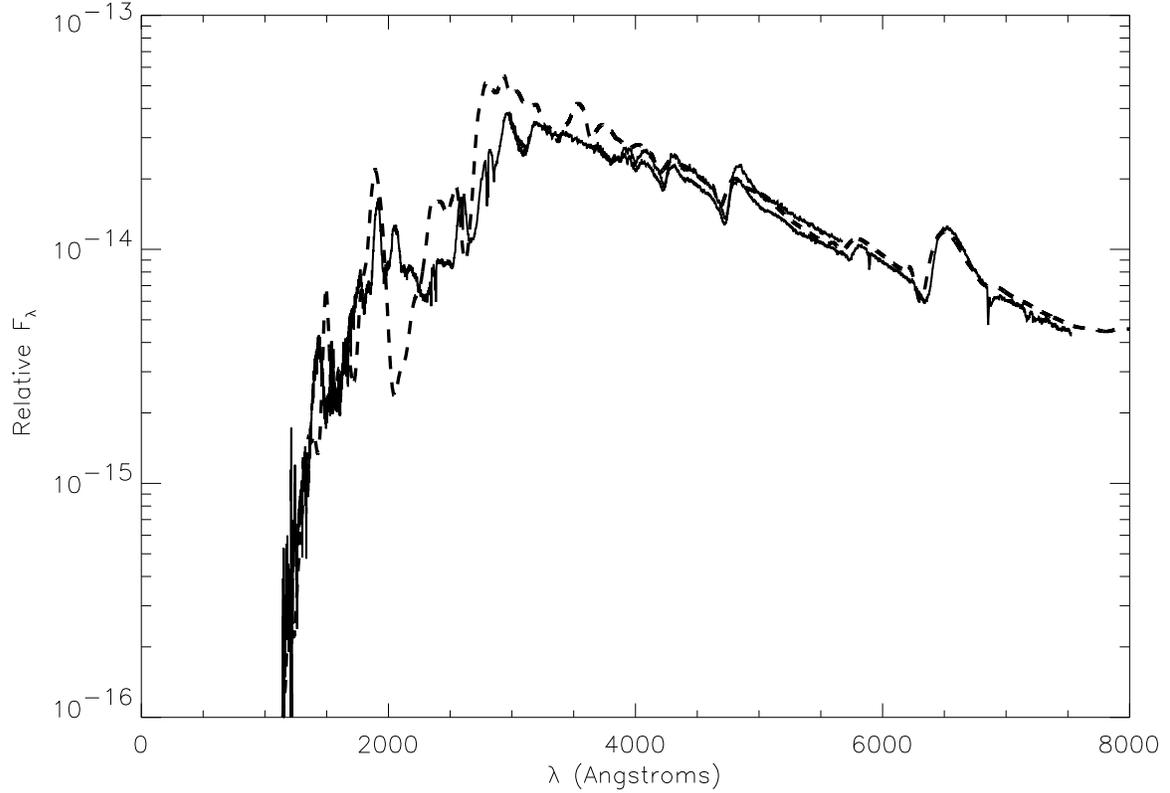}
\caption{\label{model11000} \phx\ fit (dashed line) to the UV+optical
\emph{HST} 
spectrum obtained on Nov. 5. The model temperature $\Teff=9000$~K,
$n=9$, and the compositions have He and N enhanced. Also displayed is
the optical spectrum (redward of 3600~\ang) obtained at the FLW
Observatory on Nov 4, which strengthens the case for enhanced
helium. The observed spectra have been 
de-reddened using $E(B-V)=0.10$.}
\end{center}
\end{figure}

The Galactic extinction to NGC 1637 corresponds to a color excess of
$E(B-V) =0.03$~mag, but the observed spectrum of SN~1999em shows a
distinct NaD interstellar absorption line that may indicate additional
reddening in the parent galaxy. We have calculated a grid  of
detailed fully line-blanketed \texttt{PHOENIX} models in order to
determine the ``temperature'' of the observed spectrum. With most
supernova spectra we can trade off higher temperatures for larger
reddening within prescribed limits (such that the atmosphere does not
become so hot or cool that strong unobserved lines would
predominate). Here, with this very early, very blue spectrum, we found
that we were not able to freely exchange temperature for reddening,
so that the reddening was well determined to be $E(B-V) \approx
0.05$  and the ``temperature'' was about
11,000~K. Figure~\ref{model11211} displays a fit with $\Teff=11,000$~K
and solar compositions; the observed spectrum has been dereddened with
$E(B-V) =0.05$~mag. The abundances were taken to be solar
throughout the model atmosphere and the density was assumed to follow
a power-law with 
\( \rho \propto r^{-n} \)
and $n=7$, somewhat shallower  than in the SYNOW
fits, though we don't regard the 
difference as being too  significant, since the true density structure
is unlikely to follow an exact power-law. We also present models with
$n=9$ (see Figure~\ref{model11000}).

The fit is rather good,  and the \Teff\ cannot be significantly reduced since
the Ca II H+K line becomes very strong in the synthetic spectrum
with $\Teff=10,000$~K, but it is weak in
the observed spectrum. Also, the observed feature due to He~I
$\lambda5876$ is not reproduced in the synthetic spectrum. Hotter
models are not ruled out but at $\Teff=12,000$~K the Ca II H+K feature
becomes weaker and there is a
clear flux deficiency in the red. Normally one expects that one
can trade off reddening for temperature in the red part of the
spectrum, and this is indeed the case
within a certain range of reddening. We assume a standard
\citep{card89} reddening law with $R=3.1$. However, since the spectrum
depends strongly on the strength of the Ca~II H+K feature, we find
that $\Teff = 11,000 \pm 500$~K;  the $\Teff=12,000$~K model
requires an extinction $E(B-V) \approx 0.15$ to reproduce the red
flux, but  it then does a poor job in the blue. Thus, we find that 
$E(B-V) \approx 0.05$ and $E(B-V) < 0.15$. We show a model with
$E(B-V) = 0.10$ below.

\begin{figure} 
\begin{center}
\leavevmode
\epsscale{0.95} 
\plotone{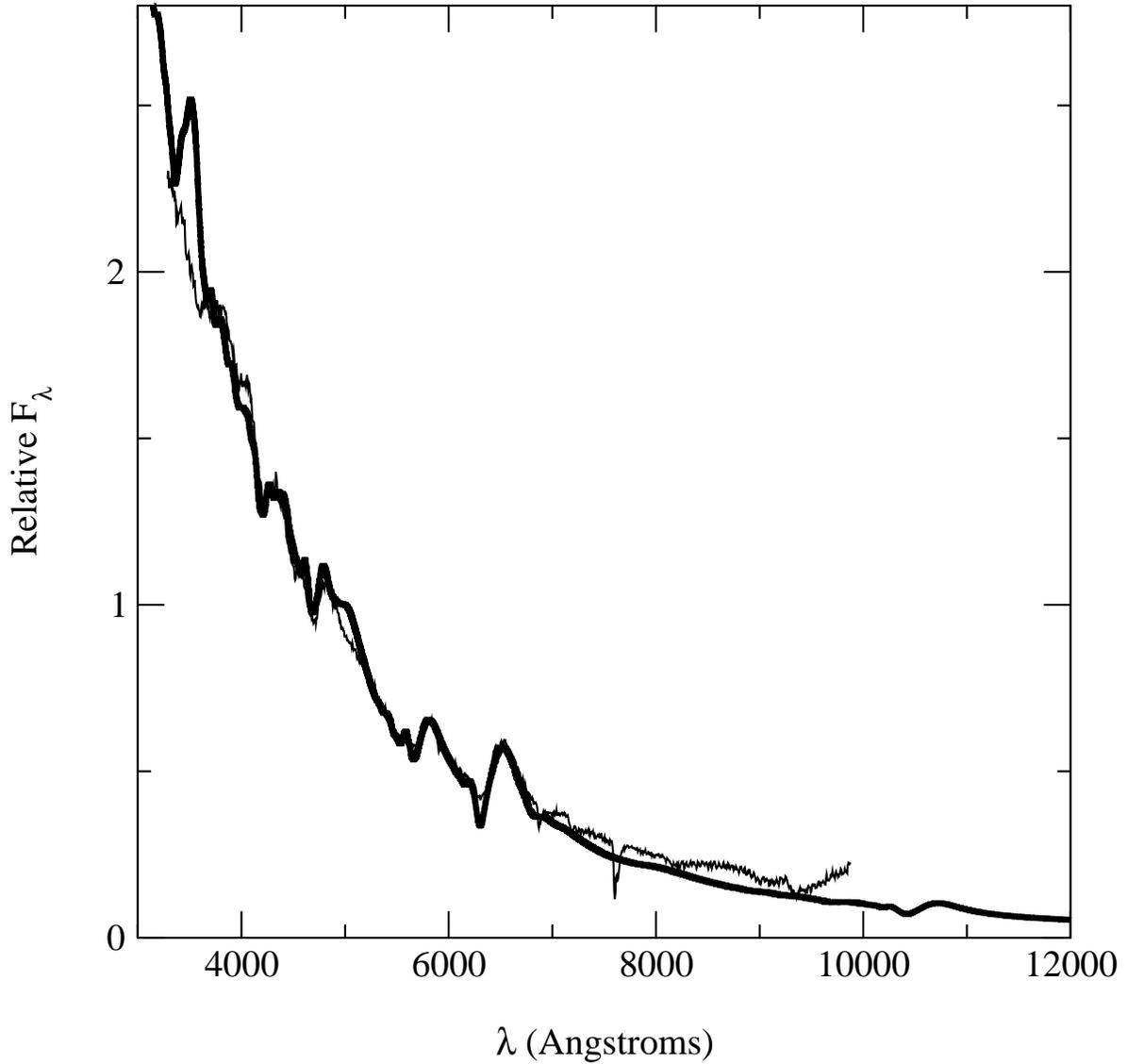}
\caption{\label{model8072} \phx\ fit (thick line) to the optical spectrum obtained
on Oct 29 at CTIO. The model temperature $\Teff=11500$~K and the
compositions are enhanced in helium and nitrogen. With this choice of
composition and temperature the He~I $\lambda 5876$ feature is well
fit, and the Ca H+K feature is too strong, indicating that the model
temperature is too low. The red flux indicates that the model
temperature is too low; thus, a slightly higher reddening may reconcile
the red and blue ends of the spectrum. The observed spectra have 
de-reddened using $E(B-V)=0.05$.}
\end{center}
\end{figure}

\begin{figure} 
\begin{center}
\leavevmode
\epsscale{0.95} 
\plotone{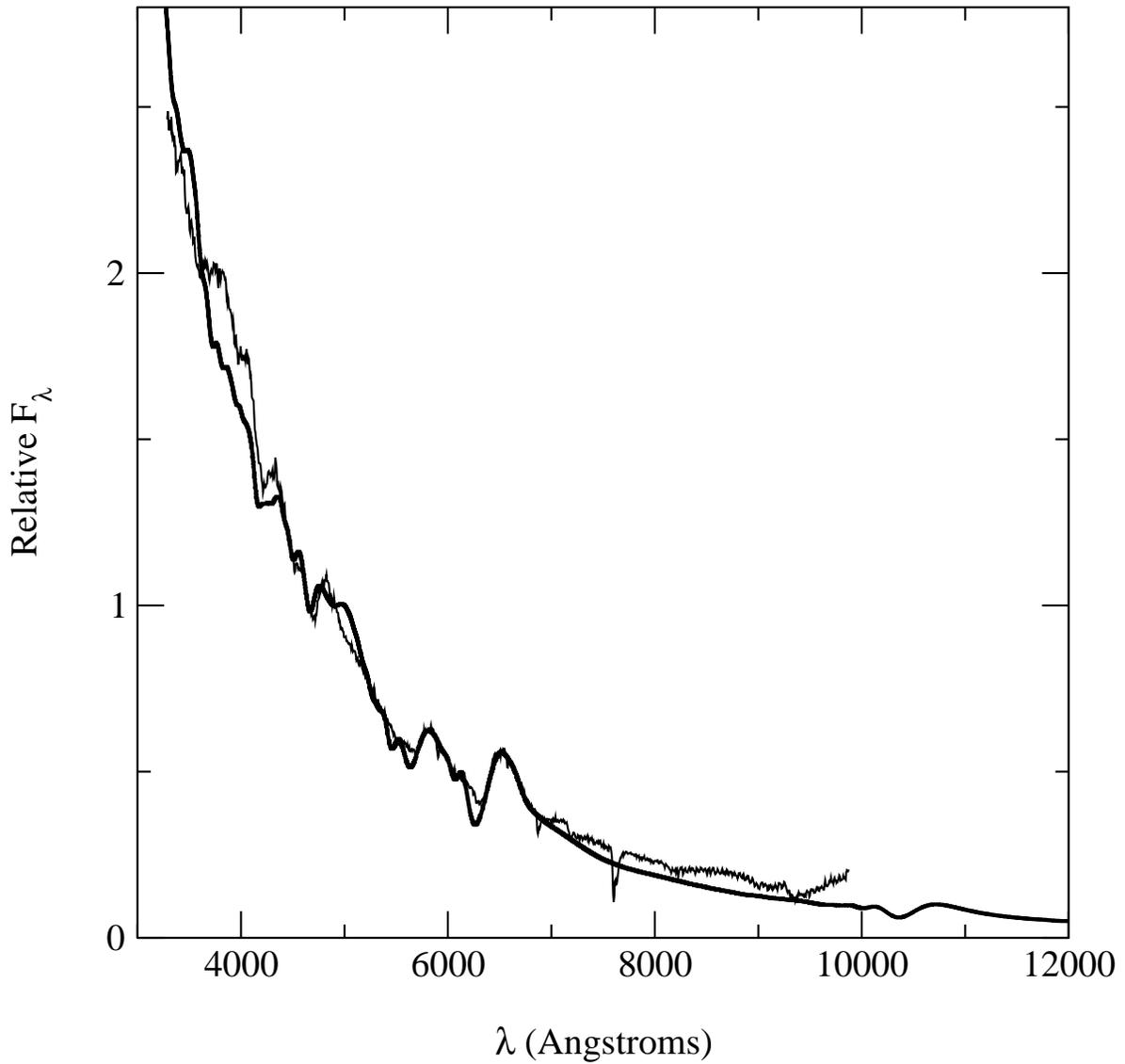}
\caption{\label{model8040}\phx\ fit (thick line) to the optical spectrum obtained
on Oct 29 at CTIO. The model temperature is $\Teff=12,000$~K and the
compositions are the enhanced ones in Figure~\ref{model8072}. The
higher temperature of this model weakens the Ca~II H+K feature seen in
Figure~\ref{model8072}, and the extinction has been taken to be
$E(B-V)=0.10$ which gives a reasonable fit to the red flux.}
\end{center}
\end{figure}

We also found that we could not
reproduce the He~I
$\lambda5876$ feature by enhancing the amount of gamma-ray
deposition. Therefore, guided by the indications of enhanced N in the
SYNOW fits, we calculated a series of spectra where we enhanced He by a
factor of 2.5 (reducing H appropriately), and we enhanced N by a factor
of 10 (reducing C+O). This fit is shown in Figure~\ref{model8072},
where now the He~I $\lambda5876$ fits well, but $\Teff =11,500$~K in
order to keep the  Ca H+K feature from becoming too strong in the
synthetic spectrum. 
Figure~\ref{model8040} with $\Teff=12,000$~K, the same compositions and
$E(B-V)=0.10$ 
shows that the effects of the Ca H+K feature can be
reduced by increasing the model temperature and increasing the
reddening modestly.
Therefore we conclude that there is strong evidence for enhanced
helium in SN 1999em [a similar result was found for SN 1987A
\citep{sn87arev,eastkir89,LF87A96,blinn87a00}]. We note at these early
times that the material is hot enough for the optical helium lines to
be excited even in LTE \citep[see][]{atlas99}, so while gamma-ray
deposition was needed in e.g. SN~1993J \citep[see][and references
therein]{b93j3}, it is not required at these early times (nor is it
precluded at later times, although we did not find it necessary for
the spectra a week later).

\subsubsection{Complicated Balmer and He~I Lines\label{complicated}}

\begin{figure} 
\begin{center}
\leavevmode
\epsscale{0.95} 
\plotone{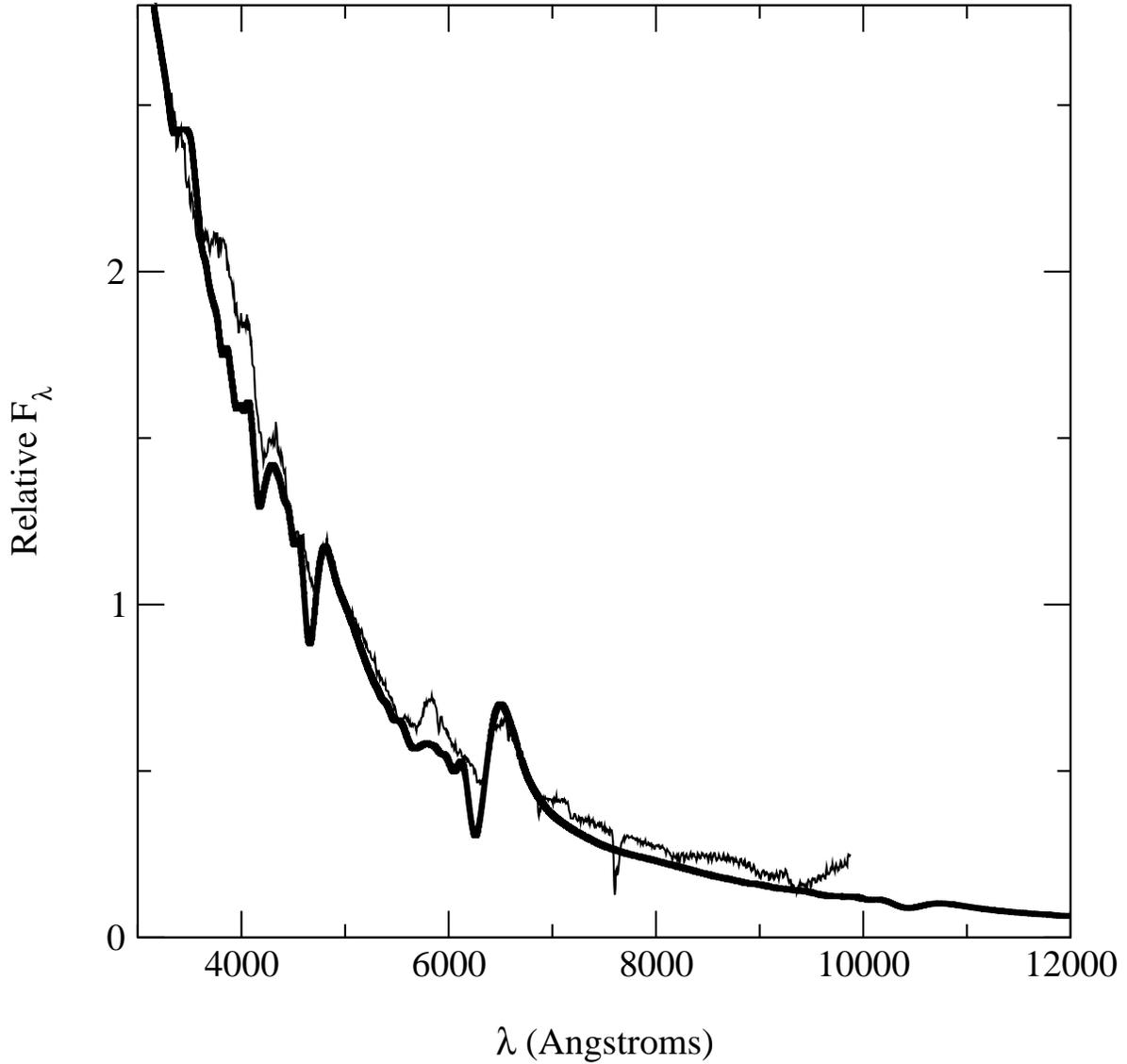}
\caption{\label{model11211} \phx\ fit (thick line) to the optical spectrum obtained
on Oct 29 at CTIO. The model temperature $\Teff=11,000$~K and the
compositions are solar. The red flux ($\lambda \ga 6500$~\ang), indicates
that the model temperature 
is about right. In this and subsequent \phx\ plots the following
species were treated in full NLTE: H~I, He~I-II, CNO I-III, Mg~II,
Ca~II, Na~I-III, Si~I-III, S~I-III, Fe~I-III, Ni~I-III, and
Co~I-III. 
}
\end{center}
\end{figure}

We attempted to confirm the evidence in the SYNOW fits that nitrogen
is also enhanced and that N~II contributes to the observed
features at $\lambda5679, \
\lambda 5029, \ \textrm{and} \lambda4623$.
 To do this we reduced the abundance of nitrogen
from its solar value by a factor of ten. This synthetic spectrum showed that 
the  $\lambda5679$ and 
$\lambda 5029$ features do decrease, but the
feature that is attributed to $\lambda4623$ in the SYNOW fits is still
present (this is not surprising, since the feature also appears in 
 Figure~\ref{model11211} with solar compositions). Guided by the LTE
line spectra presented by \citet{atlas99}, we reduced carbon by a
factor of ten from its solar value and the 4600~\ang\ feature
remained. Finally, reducing both carbon and oxygen by a factor of ten
from their solar values (leaving N fixed) still produces a feature
with if anything a better shape than in previous calculations. Thus,
with the ``usual suspects''  the source of the 4600~\ang\ feature remains
unidentified, and the evidence for enhanced nitrogen is weakened, but
it does a good job on the $\lambda5679, \
\lambda 5029,$ features.

\begin{figure} 
\begin{center}
\leavevmode
\epsscale{0.95} 
\plotone{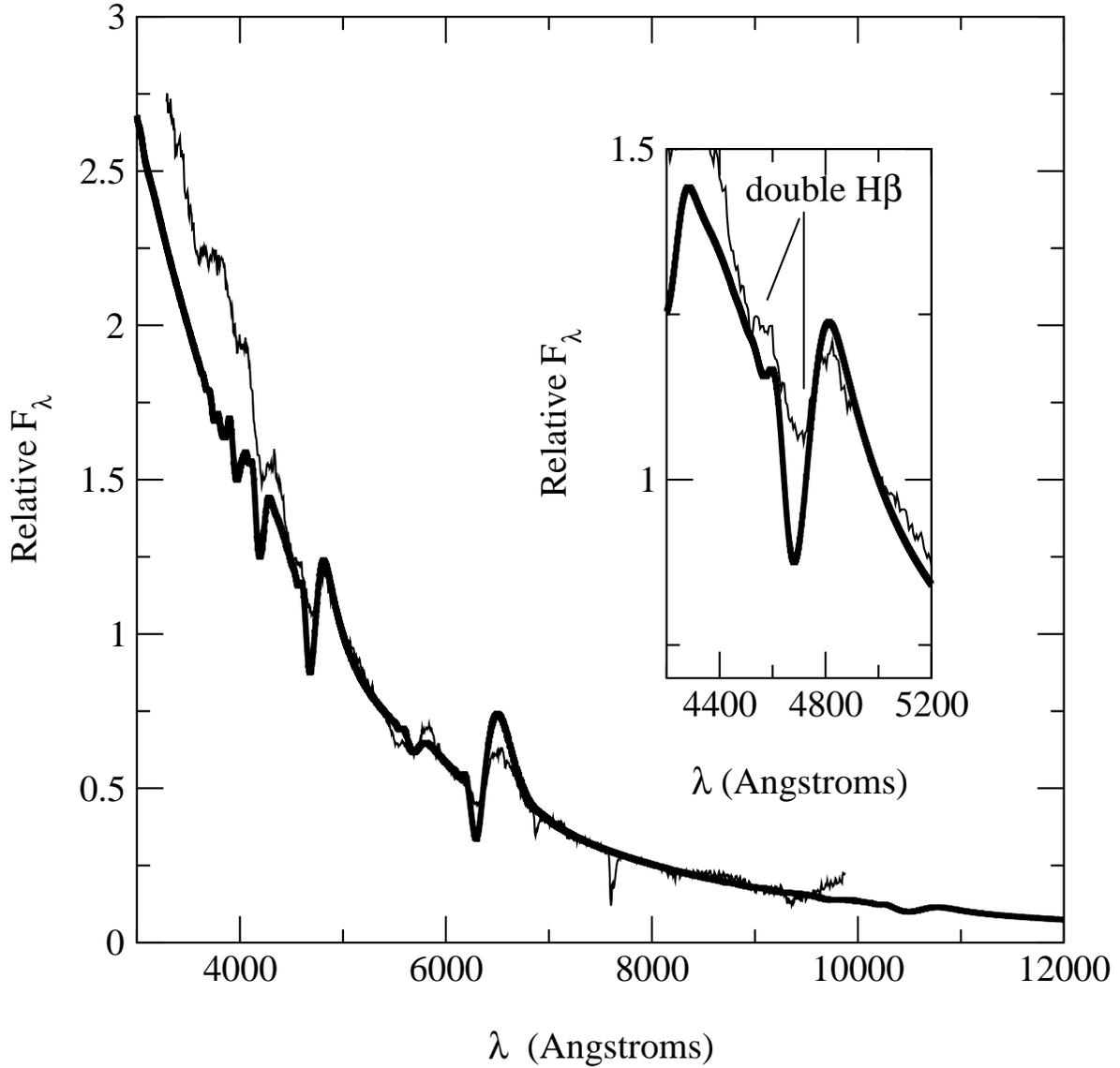}
\caption{\label{model11287}\phx\ model (thick line) with $Z=Z_\odot/100$ compared to the optical spectrum obtained on
Oct 29 at CTIO. All of the lines in the synthetic spectrum are
produced solely by hydrogen and helium. The number abundances of the
next most abundant species (C+O) are $10^{-6}$ smaller than those
of hydrogen and helium. The
reddening has been taken to be $E(B-V)=0.10$. The inset shows a close
up of H$\beta$ and it is clear that there is a shoulder in both the
observed and synthetic spectrum.
}
\end{center}
\end{figure}

 The lack of a strong candidate  for the 4600~\ang\
feature leads us to consider more exotic line identifications.
Careful examination of Figure~\ref{model11211} reveals that
even with solar compositions features appear in the synthetic spectrum
at the approximately correct wavelength positions. This led us to test
the hypothesis that the features are actually produced by ``complicated
P-Cygni'' profiles, that is the usual wide P-Cygni profile with the
peak centered near zero velocity and a second P-Cygni profile with a second
absorption minimum around 20,000~\kmps. Figure~\ref{model11287} displays
the synthetic calculation with the metalicity reduced to
$Z=Z_\odot/100$. One can be confident that all the lines in this
synthetic spectrum are produced solely by hydrogen and helium and it
is clear that each line is associated with  high-velocity
absorption, which matches the observed lines very well. (It may be a
true P-Cygni feature, with both a peak and 
a dip, but it is probably premature to decide that at this point.)
 This is due to the shallow
density gradient and high model temperature which keeps the actual
electron temperature above 8000~K all the way to the highest velocity
in the model, 37,500~\kmps. Complicated NLTE effects vary the Balmer
level populations in the mid-velocity range, producing the double
P-Cygni lines, i.e. there exist two line forming regions for hydrogen
and helium. This is the first time that we have encountered this
result and it shows that
particularly in differentially expanding flows, one cannot be certain
of line identifications until careful synthetic spectral models have
been calculated. Thus, we therefore believe that the features that we
have heretofore attributed to N~II (and which would require enhanced
N) are in actuality due to H~I and He~I, which form secondary features
at high velocity. Since this appears in our modeling without any
special parameters (i.e. Figure~\ref{model11211}), we believe this is
a true NLTE effect produced simply by the shallow density gradient.

\subsection{Nov 4--5}

\begin{figure} 
\begin{center}
\leavevmode
\epsscale{0.95} 
\plotone{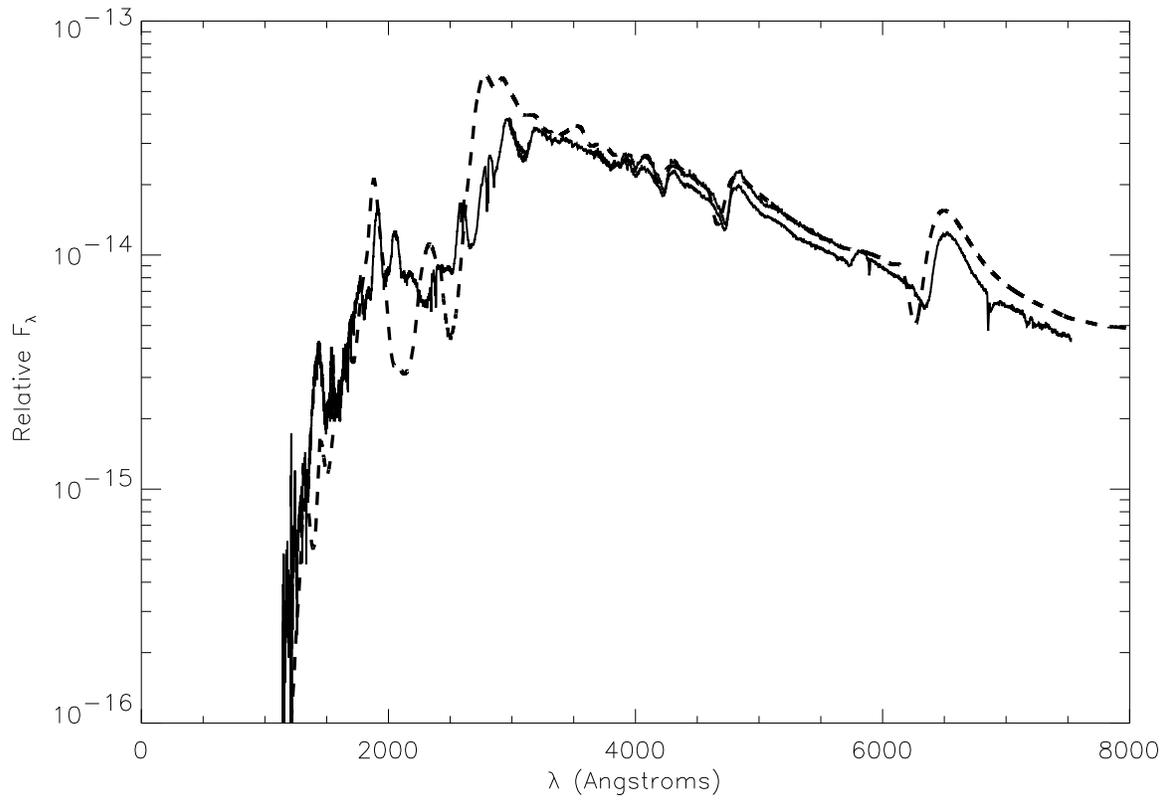}
\caption{\label{model7983} \phx\ fit (dashed line) to the UV+optical \emph{HST}
spectrum obtained on Nov. 5, with the optical spectrum obtained at the
FLWO on Nov. 4 attached (redward of 3600~\ang). The model temperature
$\Teff=8100$~K and the compositions are solar. The observed spectra
have been de-reddened by
$E(B-V)=0.05$.}
\end{center}
\end{figure}

Guided by our results in \S~\ref{phxoct29} we have computed full NLTE
models for the epoch observed with \emph{HST} using both solar and
enhanced He+N compositions (since we have not ruled out enhanced N,
we will continue to use it for expediency). Figure~\ref{model7983}
shows our best fit 
for solar compositions and $n=7$ with $\Teff=8100$~K. Figure~\ref{model11000}
displays our best fit for the enhanced He+N compositions, where we
have also found the fit is improved by setting $\Teff=9000$ and  $n=9$. Also
displayed is a ground-based optical spectrum obtained on Nov. 4, 1999
at the FLWO,
which lends support for enhanced He. All the models show unobserved
strong features in the range 2000-3500~\ang, which could be due to
iron-peak elements. Thus, it is possible that the metallicity of SN
1999em was somewhat less than solar, but we will study this question
in future work.

\section{Conclusions}

We have shown that both direct synthetic spectral fits and detailed
NLTE models do a good job of reproducing the observed optical+UV
spectra of SN~1999em. We
have shown that detailed NLTE spectral modeling of very early spectra
of Type II supernova can provide an independent estimate to the total
reddening as well as abundance information. 
Specifically we find that $E(B-V) \approx 0.05-0.10$ and $E(B-V) \la
0.15$~mag, and we find strong evidence for helium enhanced by at least a
factor of 2 over the solar value. We obtain the striking result
that observed features are likely due to ``complicated P-Cygni'' lines
of hydrogen 
and helium, and thus one needs to be especially careful with line
identifications in hot differentially expanding flows. 
That the observed complicated profiles fall right out of the
\texttt{PHOENIX} calculations is remarkable and lends strong support
to the reliablity of the modeling.

The fact that planned
supernova search programs will preferentially find supernovae very
early, combined with our results that the extinction can be determined
by modeling the very early spectra, lends promise to the use of Type II
supernovae as distance indicators, through the use of phenomenological
applications like the ``Expanding Photosphere Method''
\citep{baadeepm,kkepm,branepm,schmkireas92,esk96} or through sophisticated modeling of the
``Spectral-fitting Expanding Atmosphere Method''
\citep{b93j1,b93j2,b93j3,b93j4,b94i1}. 

\acknowledgments We thank Doug Leonard and Weidong Li for sharing
their results in advance of publication and for helpful
discussions. PHH was supported in part by the P\^ole
Scientifique de Mod\'elisation Num\'erique at ENS-Lyon. This work was
supported in part by NSF grants 
AST-9731450, AST-9417102, and AST-9417213; by NASA grant NAG5-3505,
and an
IBM SUR grant to the University of Oklahoma; by NSF grant AST-9720704,
NASA ATP grant NAG 5-8425 and LTSA grant NAG 5-3619 to the University
of Georgia; and by NASA GO-8243 and GO-8648 to the SINS group from the Space
Telescope Science Institute, which is operated by AURA, Inc.~under NASA
contract NAS 5--26555.  Some of the calculations presented in this
paper were performed at the San Diego Supercomputer Center (SDSC),
supported by the NSF, and at the National Energy Research
Supercomputer Center (NERSC), supported by the U.S. DOE. We thank both
these institutions for a generous allocation of computer time.  This
research has made use of the NASA/IPAC Extragalactic Database (NED)
which is operated by the Jet Propulsion Laboratory, California
Institute of Technology, under contract with the National Aeronautics
and Space Administration.

\bibliography{refs,baron,sn1bc,sn1a,snii,sn87a,rte,crossrefs}

\clearpage

\begin{figure} 
\begin{center}
\leavevmode
\epsscale{0.95} 
\plotone{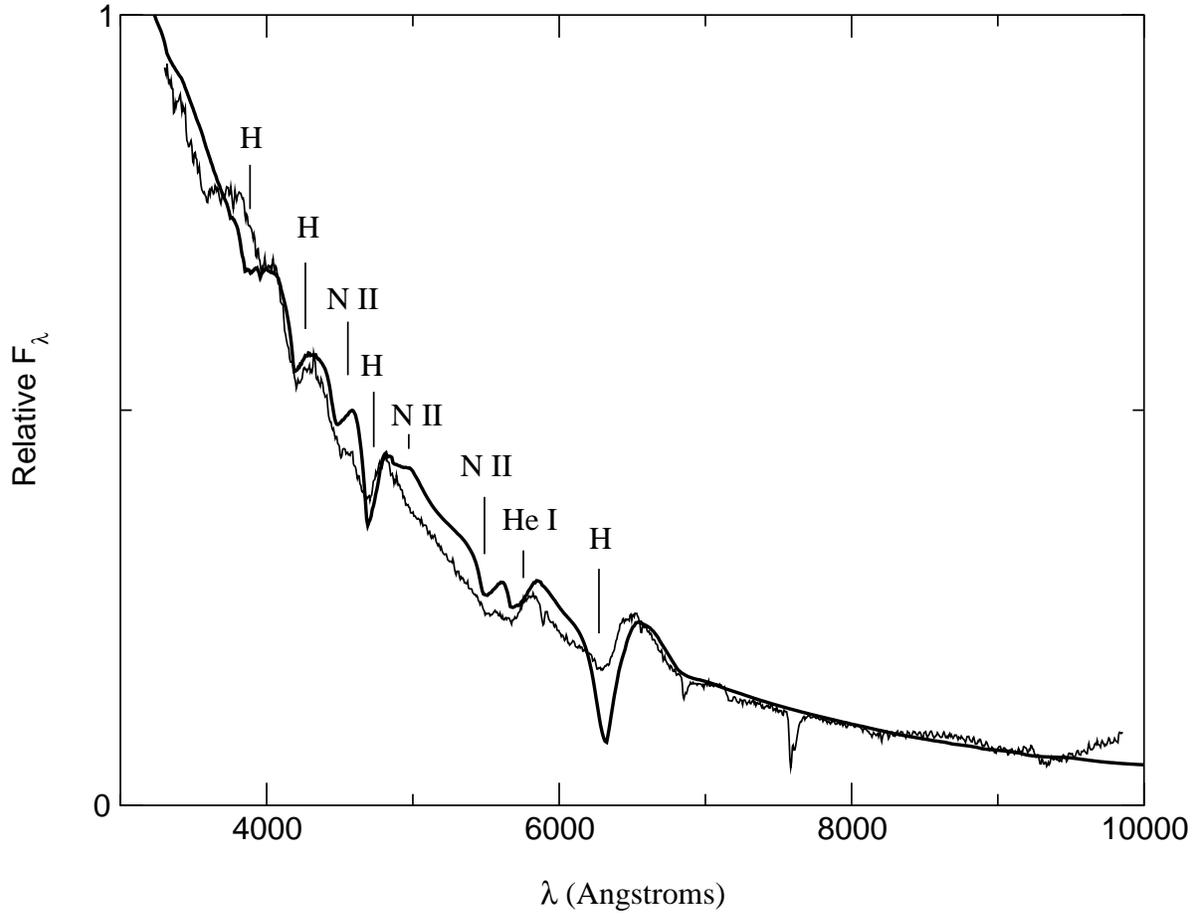}
\caption{\label{syn_oct29} SYNOW fit to Optical Oct.~29
spectrum. Observed spectra have been de-redshifted by 717~\kmps\
in this and subsequent figures.
The observed spectrum has been de-reddened using $E(B-V)=0.05$.}
\end{center}
\end{figure}

\clearpage

\end{document}